\documentclass[notitlepage,12 pt,twoside,aps,prd,amsmath,amssymb,
tightenlines,showpacs,showkeys,eqsecnum]{revtex4-1}

\usepackage[pagebackref]{hyperref}
\usepackage{graphicx}
\usepackage{mathptmx}
\usepackage{bm}
\usepackage{amssymb}
\usepackage{amsmath}
\newcommand {\cG}{\cal G}
\newcommand {\cD}{\cal D}
\newcommand {\bg}{\bar \gamma}
\newcommand {\G}{\Omega}
\newcommand {\bp}{\bar \psi}

\def\myfigure #1#2#3#4
{\begin{figure}[ht]\begin{center}
\includegraphics[width=#2 \textwidth]{#1.eps}
\parbox[t]{#4\textwidth}{\caption{#3}\label{#1}}
\end{center}\end{figure}}

\def \myfigures #1#2#3#4#5#6#7#8
{\begin{figure}[ht]
    \begin{center}
        \includegraphics[width=#2 \textwidth]{#1.eps}
        \hfill
        \includegraphics[width=#6 \textwidth]{#5.eps}
        \parbox[t]{#4\textwidth}{\caption {#3}\label{#1}}
        \hfill
        \parbox[t]{#8\textwidth}{\caption {#7}\label{#5}}
    \end{center}
\end{figure} }

\begin{document}

\baselineskip -24pt
\title{Time-dependent Spinor field in a static cylindrically symmetric space-time}
\author{Bijan Saha}
\affiliation{Laboratory of Information Technologies\\
Joint Institute for Nuclear Research\\
141980 Dubna, Moscow region, Russia\\ and\\
Peoples' Friendship University (RUDN University)\\
Moscow, Russia} \email{bijan@jinr.ru}
\homepage{http://spinor.bijansaha.ru}

\hskip 1 cm

\begin{abstract}
Within the scope of a static cylindrically symmetric space-time we
study the behavior of a nonlinear spinor field that depends on time
and radial coordinates. It is found that the presence of nontrivial
non-diagonal components of the energy-momentum tensor (EMT) imposes
some restrictions on both the metric functions and the spinor field.
While for the time independent spinor field there occur three way
restrictions resembling the Bianchi type-I Universe, in this case
things become more complicated.
\end{abstract}



\maketitle

\bigskip

\section{Introduction}

From the physical point of view the spherically symmetric solutions
to the Einstein equations with nontrivial energy-momentum tensor
present major interest, since this class of space-time is compatible
with asymptotic flatness. Another class of asymptotically flat
space-time is the axially symmetric ones. Due to mathematical
difficulties related to axially symmetric space-time many authors
consider cylindrically symmetric space-time as a preliminary step,
though the later one is not asymptotically flat \cite{fjalborg}. The
cylindrical symmetric solutions localized in the vicinity of the
axis of symmetry, such as vortices was studied in \cite{Neilsen}. As
it was shown in \cite{Terlet} even the simplest nonlinear equation
of complex scalar field can provide with not only particle-like
solutions, but also string-like solutions which was described by a
cylindrical symmetric space-time. These solutions may describe
realistic objects such as super-conducting threads - fluxons
\cite{Abrikos} or light beams \cite{Zakharov} and serve as
reasonable approximation to objects with a toroidal structure where
large toroid radius is approximately replaced by a closed string
segment \cite{Vega}.

Within the framework of a static cylindrically symmetric space-time
Levi-Civita discovered a class of solutions of Einstein field
equations in vacuum \cite{Levi-Civita}. Further this was developed
by many authors. An interacting system of scalar, electromagnetic
and gravitational fields within the scope of cylindrical symmetric
space-time was studied in \cite{Shikin1995,sahaijtp97}. The
corresponding equations were exactly solved and soliton-like
solutions as a whole and droplet-like solutions in particular, were
obtained. There has been increasing interest in the study of
neutrino stars in astrophysics. In doing so static cylindrically
symmetric solutions to the Einstein field equation with elastic
matter was obtained in \cite{brito}. The possibility of forming
anisotropic compact stars from cosmological constant as one of the
competent candidates of dark energy with cylindrical symmetry was
discussed in \cite{Abbas}. A comprehensive review of cylindrical
symmetric systems in GR can be found in \cite{bron-cyl}.

In most cases perfect fluid and scalar fields are considered as the
the source of the corresponding gravitational fields. In recent
years some alternatives to those are also being exploited. Since the
spinor field is very sensitive to the gravitational one and can
generate different kinds of fluids and dark energy, many authors
consider this field in cosmological problems
\cite{SahaPRD2001,SahaPRD2006,SahaECAYA2018}.

Lately spinor field is being used in various fields that includes
gravity as well. The coupled Einstein-Dirac equations for a static,
spherically symmetric system of two fermions in a singlet spinor
state are derived and soliton-like solutions to these equations were
found by Finster et al \cite{Finster}. A comparative analysis of
three different types of solitonic solutions of GR-matter systems,
which can be interpreted as explicit realizations of Wheeler's geon
concept for matter fields of spin 0, 1 and 1/2, respectively, was
provided in \cite{Herdeiro}. The authors identified the common
conditions that allow these solutions. A comparison of spherically
symmetric geons for the spin 0,1/2 and 1, emphasizing the
mathematical similarities and clarifying the physical differences,
particularly between the bosonic and fermionic cases, is also
presented. Beside these Einstein-Dirac system was studied in
\cite{Leith1,Leith2}.

In a recent paper \cite{universe} we have studied the behavior of a
nonlinear spinor field within the scope of a static
cylindrical-symmetric space-time. In that paper we have considered
the spinor field that depends on the radial coordinate only.

In this report we will extend that study and consider the spinor
field that depends on time and radial coordinates within the
framework of a static cylindrical-symmetric space-time.

\section{Basic equations and their solutions}

The spinor field Lagrangian we take in the form \cite{SahaPRD2001}

\begin{equation}
L_{\rm sp} = \frac{\imath}{2} \left[\bp \gamma^{\mu} \nabla_{\mu}
\psi- \nabla_{\mu} \bar \psi \gamma^{\mu} \psi \right] - m_{\rm sp}
\bp \psi - \lambda F(K), \label{lspin}
\end{equation}
with the nonlinear term being some arbitrary function of the
invariant $K = \{I,\,J,\,I+J,\, I - J\}$ i.e., $F = F(K)$, where $I
= S^2 = (\bp \psi)^2$ and $J = P^2 = (\imath \bp \gamma^5 \psi)^2$.

The spinor field equations corresponding to the Lagrangian
\eqref{lspin} are

\begin{subequations}
\label{speq}
\begin{eqnarray}
\imath\gamma^\mu \nabla_\mu \psi -\left( m_{\rm sp} + {\cD}\right)
\psi -
 \imath {\cG} \gamma^5 \psi &=&0, \label{speq1} \\
\imath \nabla_\mu \bp \gamma^\mu + \left( m_{\rm sp} +
{\cD}\right)\bp + \imath {\cG} \bp \gamma^5 &=& 0, \label{speq2}
\end{eqnarray}
\end{subequations}
where we denote ${\cD} = 2 \lambda S F_K K_I$ and ${\cG} = 2 \lambda
P F_K K_J$, with $F_K = dF/dK$, $K_I = dK/dI$ and $K_J = dK/dJ.$ In
view of \eqref{speq} it can be shown that
\begin{equation}
L_{\rm sp} = \lambda\left( 2 K F_K - F\right). \label{LspinF}
\end{equation}

The energy-momentum tensor (EMT) of the spinor field is obtained
from
\begin{equation}
T_{\mu}^{\,\,\,\rho}=\frac{\imath}{4} g^{\rho\nu} \biggl(\bp
\gamma_\mu \nabla_\nu \psi + \bp \gamma_\nu \nabla_\mu \psi -
\nabla_\mu \bar \psi \gamma_\nu \psi - \nabla_\nu \bp \gamma_\mu
\psi \biggr) \,- \delta_{\mu}^{\rho} L_{\rm sp}. \label{temsp}
\end{equation}

Taking into account that the covariant derivative of the spinor
field $\nabla_{\mu}$ has the form
\begin{equation}
\nabla_\nu \psi = \partial_\nu \psi - \G_\mu \psi, \quad \nabla_\nu
\bp = \partial_\nu \bp + \G_\mu \bp,
\end{equation}
\eqref{temsp} can be written as

\begin{eqnarray}
T_{\mu}^{\,\,\,\rho}&=&\frac{\imath}{4} g^{\rho\nu} \biggl(\bp
\gamma_\mu
\partial_\nu \psi + \bp \gamma_\nu \partial_\mu \psi -
\partial_\mu \bar \psi \gamma_\nu \psi - \partial_\nu \bp \gamma_\mu
\psi \biggr)\nonumber\\
& - &\frac{\imath}{4} g^{\rho\nu} \bp \biggl(\gamma_\mu \G_\nu +
\G_\nu \gamma_\mu + \gamma_\nu \G_\mu + \G_\mu \gamma_\nu\biggr)\psi
 \,- \delta_{\mu}^{\rho} L_{\rm sp}. \label{temsp0}
\end{eqnarray}

Here $\G_\nu$ is the spinor affine connection constructed from the
metric function.

Let us consider the static cylindrically symmetric space-time given
by the metric element \cite{brito}
\begin{equation}
ds^2 = e^{2\gamma} dt^2 - e^{2 \alpha} du^2 - e^{2 \beta} d\phi^2 -
e^{2 \mu} dz^2,  \label{cylin}
\end{equation}
where the spacetime coordinates are $x^\mu = \{t,\,u,\,\phi,\,z\}$
and $\gamma,\, \alpha,\, \beta$ and $\mu$ are $C^2$ functions of $u$
only.

In analogy to standard spherically symmetric spacetime, it is
customary to define the radial coordinate $r$ in such a way that
co-efficient of $d\phi^2$ is equal to $r^2$, i.e. $e^{2 \beta} =
r^2$. This transformation is called tangential gauge \cite{Abbas}.

The spinor affine connection $\G_\mu$ is defined as
\begin{equation}
\G_\mu = \frac{1}{4} g_{\rho \sigma} \left(\partial_\mu e^{(b)}_\tau
e^\rho_{(b)} - \Gamma_{\mu \tau}^\rho\right)\gamma^\sigma
\gamma^\tau, \label{SACDef}
\end{equation}
where $e^\rho_{(b)}$ and $e^{(b)}_\tau$ are the tetrad vectors such
that  $e^\rho_{(b)} e^{(b)}_\tau = \delta^\rho_\tau$ and
$e^\rho_{(b)} e^{(a)}_\rho = \delta^a_b$. From \eqref{SACDef} we
find the spinor affine connection corresponding to the metric
\eqref{cylin}:

\begin{align}
\G_0 &= -\frac{1}{2} e^{\gamma - \alpha}\, \gamma^\prime\, \bg^0
\bg^1, \quad \G_1 = 0, \quad \G_2 = \frac{1}{2} e^{\beta - \alpha}\,
\beta^\prime\, \bg^2 \bg^1, \quad \G_3 &= \frac{1}{2} e^{\mu -
\alpha}\, \mu^\prime\, \bg^3 \bg^1, \label{SAC}
\end{align}
where prime ($^\prime$) denotes differentiation with respect to $u$.

Let us now consider the case when the spinor field depends on $x^0 =
t$ and $x^1 = u$ such that $\psi = e^{-\imath\omega t} \psi (u)$ and
$\bp = e^{\imath\omega t} \bp(u)$.

The spinor field equations \eqref{speq} in this case read

\begin{subequations}
\label{speqcs1n}
\begin{eqnarray}
\psi^\prime + \frac{1}{2} \tau^\prime \psi - \imath \omega e^{\alpha
- \gamma}\bg^0 \bg^1 \psi - \imath {\cD}_1 e^{\alpha} \bg^1\psi -
{\cG} e^{\alpha} \bg^5 \bg^1 \psi &=&0, \label{speq1cs1n} \\
\bp^\prime + \frac{1}{2} \tau^\prime \bp  - \imath \omega e^{\alpha
- \gamma} \bp \bg^0 \bg^1 + \imath {\cD}_1 e^{\alpha} \bp \bg^1 -
{\cG} e^{\alpha} \bp \bg^5 \bg^1 &=&0, \label{speq2cs1n}
\end{eqnarray}
\end{subequations}

where we denote ${\cD}_1 := \left( m_{\rm sp} + {\cD}\right)$ and

\begin{equation}
\tau  = \gamma + \beta + \mu. \label{tau}
\end{equation}

Taking into account that $S = \bp \psi$, $P = \imath \bp \bg^5 \psi$
and denoting  ${\bar v}^\mu =  \bp \bg^\mu \psi$, ${\bar A}^\mu =
\imath \bp \bg^5 \bg^\mu \psi$ and ${\bar Q}^{\mu\nu} = \bp \bg^\mu
\bg^\nu \psi$ from the foregoing equations one finds the following
relations

\begin{subequations}
\label{speqcs1nm}
\begin{eqnarray}
\bp \bg^0 \psi^\prime - \bp^\prime \bg^0 \psi - 2\imath \omega
e^{\alpha - \gamma} {\bar v^1} &=& 0, \label{v0m} \\
\bp \bg^1 \psi^\prime - \bp^\prime \bg^1 \psi - 2\imath \omega
e^{\alpha - \gamma} {\bar v^0} + 2 \imath e^\alpha {\cD}_1 S + 2 \imath e^\alpha {\cG} P &=& 0, \label{v1m} \\
\bp \bg^2 \psi^\prime - \bp^\prime \bg^2 \psi  &=& 0, \label{v2m} \\
\bp \bg^3 \psi^\prime - \bp^\prime \bg^3 \psi  &=& 0, \label{v3m}
\end{eqnarray}
\end{subequations}

and

\begin{subequations}
\label{speqcs1n0}
\begin{eqnarray}
S_0^\prime - 2 \imath \omega e^{\alpha - \gamma} \bar
Q^{01}_0  - 2 {\cG} e^\alpha \bar A^1_0 &=& 0. \label{S0}\\
P_0^\prime - 2 \imath \omega e^{\alpha - \gamma} \bar
Q^{23}_0 - 2 {\cD}_1 e^\alpha \bar A^1_0 &=& 0. \label{P0}\\
\bar v^{0\prime}_0 - 2 \imath {\cD}_1 e^\alpha \bar Q^{01}_0 - 2
\imath {\cG} e^\alpha \bar Q^{23}_0 &=& 0, \label{v00} \\
\bar v^{1\prime}_0 &=& 0, \label{v10} \\
\bar v^{2\prime}_0 + 2 \omega e^{\alpha - \gamma} \bar A^3_0 + 2
\imath {\cD}_1 e^\alpha \bar Q^{12}_0 - 2 \imath {\cG} e^\alpha \bar
Q^{03}_0 &=& 0, \label{v20} \\
\bar v^{3\prime}_0 + 2 \omega e^{\alpha - \gamma} \bar A^2_0 - 2
\imath {\cD}_1 e^\alpha \bar Q^{13}_0 + 2 \imath {\cG} e^\alpha \bar
Q^{02}_0 &=& 0, \label{v30} \\
\bar A^{0\prime}_0 &=& 0, \label{A00} \\
\bar A^{1\prime}_0 + 2 {\cD}_1 e^{\alpha}P_0 + 2
{\cG} e^\alpha S_0  &=& 0, \label{A10} \\
\bar A^{2\prime}_0 - 2  \omega e^{\alpha
- \gamma} \bar v^3_0 &=& 0, \label{A20} \\
\bar A^{3\prime}_0 + 2 \omega e^{\alpha
- \gamma} \bar v^2_0 &=& 0, \label{A30}\\
\bar Q^{01 \prime}_0 - 2 \imath \omega e^{\alpha - \gamma} S_0 + 2
\imath {\cD}_1 e^\alpha \bar v^0_0 &=& 0, \label{q010}\\
\bar Q^{02 \prime}_0  -  2 \imath {\cG}
e^\alpha \bar v^3_0 &=& 0, \label{q020}\\
\bar Q^{03 \prime}_0 +  2 \imath {\cG}
e^\alpha \bar v^2_0 &=& 0, \label{q030}\\
\bar Q^{12 \prime}_0 -  2 \imath {\cD}_1
e^\alpha \bar v^2_0 &=& 0, \label{q120}\\
\bar Q^{13 \prime}_0 -  2 \imath {\cD}_1
e^\alpha \bar v^3_0 &=& 0, \label{q130}\\
\bar Q^{23 \prime}_0 - 2 \imath \omega e^{\alpha - \gamma} P_0 + 2
\imath {\cG} e^\alpha \bar v^0_0 &=& 0, \label{q230}
\end{eqnarray}
\end{subequations}
where we denote $S_0 = S e^\tau$, $P_0 = P e^\tau$, $\bar v^\mu_0 =
\bar v^\mu e^\tau$, $\bar A^\mu_0 = \bar A^\mu e^\tau$ and $\bar
Q^{\mu\nu}_0 = \bar Q^{\mu\nu} e^\tau$.

The foregoing system gives the following first integrals

\begin{subequations}
\label{firstint}
\begin{eqnarray}
\bar A_0^0 &= & C_1, \label{firstint1}\\
\bar v_0^1 &= & C_2, \label{firstint2}\\
\left(S_0\right)^2 + \left(P_0\right)^2 + \left(\bar A_0^1\right)^2
- \left(\bar v_0^0\right)^2 - \left(\bar Q_0^{01}\right)^2 -
\left(\bar Q_0^{23}\right)^2 &=& C_3^2,  \label{firstint3}\\
\left(\bar v_0^2\right)^2 + \left(\bar v_0^3\right)^2 + \left(\bar
A_0^2\right)^2 + \left(\bar A_0^3\right)^2 + \left(\bar
Q_0^{02}\right)^2 + \left(\bar Q_0^{03}\right)^2 + \left(\bar
Q_0^{12}\right)^2 - \left(\bar Q_0^{13}\right)^2 &=& C_4^2.
\label{firstint4}
\end{eqnarray}
\end{subequations}
where $C_i$ are the integration constants. From \eqref{firstint} it
is clear that each term in these equalities, i.e. $S_0, P_0,\, \bar
A^\mu_0,\, \bar v_0^\mu,\, \bar Q_0^{\mu\nu}$ are constants.

Finally we have the following nontrivial components of EMT

\begin{subequations}
\label{EMTgencs2}
\begin{align}
T_0^0 &= \omega  e^{-\gamma} \bar v^0 + \lambda \left(F - 2 K F_K\right), \label{00cs2}\\
T_1^1 &= m_{\rm sp} S + \lambda F - \omega e^{- \gamma } \bar v^0, \label{11cs2}\\
T_2^2 &= \lambda \left(F - 2 K F_K\right), \label{22cs2}\\
T_3^3 &= \lambda \left(F - 2 K F_K\right), \label{33cs2}\\
T_1^0 &= -\omega e^{\alpha - 2 \gamma } \bar v^1, \label{01cs2}\\
T_2^0 &= -\frac{\omega}{2} e^{\beta - 2 \gamma } \bar v^2 -
\frac{1}{4} \left(\gamma^\prime - \beta^\prime\right) e^{\beta -
\gamma - \alpha } \bar A^3,\label{02cs2}\\
T_3^0 &= -\frac{\omega}{2} e^{\mu - 2 \gamma } \bar v^3 -
\frac{1}{4} \left(\gamma^\prime - \mu^\prime\right) e^{\mu -
\gamma - \alpha } \bar A^2, \label{03cs2}\\
T_3^2 &= \frac{1}{4} \left(\beta^\prime - \mu^\prime\right) e^{\mu -
\beta - \alpha } \bar A^0. \label{23cs2}
\end{align}
\end{subequations}

From \eqref{EMTgencs2} one sees, that $T_0^0 \ne T_2^2$ and

\begin{subequations}
\label{EMTgencs2}
\begin{align}
T_0^0 + T_1^1 &= m_{\rm sp} S + 2\lambda \left(F - K F_K\right), \label{00+11}\\
T_0^0 - T_1^1 &= 2\omega  e^{-\gamma} \bar v^0 - m_{\rm sp} S - 2
\lambda K F_K. \label{00-11}
\end{align}
\end{subequations}

Before dealing with the Einstein system of equations let us remark
that the Bianchi identity, i.e., $G^\nu_{\mu;\nu} = 0$ leads to
$T^\nu_{\mu;\nu} = 0$. Indeed, on account of \eqref{EMTgencs2} we
find

\begin{eqnarray}
T^\nu_{\mu;\nu} &=& T^\nu_{\mu,\nu} + \Gamma^{\nu}_{\rho \nu}
T^\rho_\mu - \Gamma^\rho_{\mu\nu} T^\nu_\rho \nonumber\\
&=& m_{\rm sp} e^{-\tau} \frac{\partial}{\partial u} \left(S
e^\tau\right) + \lambda F_K e^{-2 \tau} \frac{\partial}{\partial u}
\left(K e^{2\tau}\right) - \omega e^{-\gamma - \tau}
\frac{\partial}{\partial u} \left(\bar v^0 e^\tau\right) = 0.
\label{BIden}
\end{eqnarray}

As it was mentioned earlier $S e^\tau = S_0$,\, $ K e^{2\tau} = K_0$
and $\bar v^0 e^\tau = \bar v^0_0$ are constants, hence the
conservation law $T^\nu_{\mu;\nu} = 0$ is fulfilled identically.

Since, for the metric \eqref{cylin} the Einstein tensor has only
diagonal components, so let us first consider the diagonal equations
of Einstein system

\begin{subequations}
\begin{align}
e^{-2\alpha} \left[\gamma^\prime \beta^\prime + \beta^\prime
\mu^\prime + \mu^\prime \gamma^\prime\right] &= m_{\rm sp} S_0
e^{-\tau} + \lambda F - \omega e^{- \gamma - \tau} \bar v^0_0,
\label{EE11}\\
e^{-2\alpha} \left[\gamma^{\prime\prime} + \mu^{\prime\prime} +
\gamma^{\prime 2} + \mu^{\prime 2} + \gamma^{\prime} \mu^{\prime} -
\alpha^{\prime}\gamma^{\prime} - \alpha^{\prime}\mu^{\prime}\right]
&=   \lambda \left( F - 2 K F_K \right), \label{EE22}\\
e^{-2\alpha} \left[\gamma^{\prime\prime} + \beta^{\prime\prime} +
\gamma^{\prime 2} + \beta^{\prime 2} + \gamma^{\prime}
\beta^{\prime} - \alpha^{\prime}\gamma^{\prime} -
\alpha^{\prime}\beta^{\prime}\right] &=   \lambda \left( F - 2 K F_K
\right), \label{EE33}\\
e^{-2\alpha} \left[\beta^{\prime\prime} + \mu^{\prime\prime} +
\beta^{\prime 2} + \mu^{\prime 2} + \beta^{\prime} \mu^{\prime} -
\alpha^{\prime}\beta^{\prime} - \alpha^{\prime}\mu^{\prime}\right]
&=  \omega  e^{-\gamma - \tau} \bar v^0_0 + \lambda \left(F - 2 K
F_K\right). \label{EE00}
\end{align}
\end{subequations}

Subtraction of \eqref{EE22} from \eqref{EE00} and \eqref{EE33} from
\eqref{EE00} yields, respectively

\begin{subequations}
\begin{align}
\beta^{\prime \prime} - \gamma^{\prime \prime} + \beta^{\prime 2} -
\gamma^{\prime 2} + \mu^\prime \left(\beta^\prime -
\gamma^\prime\right)- \alpha^\prime \left(\beta^\prime -
\gamma^\prime\right) &= \omega e^{2 \alpha - \gamma - \tau}\, \bar
v_0^0,
\label{00-22}\\
\mu^{\prime \prime} - \gamma^{\prime \prime} + \mu^{\prime 2} -
\gamma^{\prime 2} + \beta^\prime \left(\mu^\prime -
\gamma^\prime\right)- \alpha^\prime \left(\mu^\prime -
\gamma^\prime\right) &= \omega e^{2 \alpha - \gamma - \tau}\,\bar
v_0^0. \label{00-33}
\end{align}
\end{subequations}

For the non-diagonal components of Einstein equations we have

\begin{subequations}
\begin{align}
0 &= -\omega e^{\alpha - 2 \gamma - \tau} \bar v^1_0, \label{EE10}\\
0 &= -\frac{\omega}{2} e^{\beta - 2 \gamma - \tau} \bar v^2_0 -
\frac{1}{4} \left(\gamma^\prime - \beta^\prime\right) e^{\beta -
\gamma - \alpha - \tau} \bar A^3_0,\label{EE20}\\
0 &= -\frac{\omega}{2} e^{\mu - 2 \gamma - \tau} \bar v^3_0 -
\frac{1}{4} \left(\gamma^\prime - \mu^\prime\right) e^{\mu -
\gamma - \alpha - \tau} \bar A^2_0, \label{EE30}\\
0 &= \frac{1}{4} \left(\beta^\prime - \mu^\prime\right) e^{\mu -
\beta - \alpha - \tau} \bar A^0_0. \label{EE23}
\end{align}
\end{subequations}

From \eqref{EE20} and \eqref{EE30} we have
\begin{eqnarray}
\beta^\prime - \gamma^\prime &=& 2 \omega
c_1\,e^{\alpha - \gamma}, \label{bet0a}\\
\mu^\prime - \gamma^\prime &=& 2 \omega  c_2\,e^{\alpha -
\gamma},\label{mu0a}
\end{eqnarray}

where we denote $c_1 = \bar v^2_0/\bar A_0^3$ and $c_2 = \bar
v^3_0/\bar A_0^2$.

In view of \eqref{bet0a} and \eqref{mu0a} the eqns. \eqref{00-22}
and \eqref{00-33} can be written as

\begin{subequations}
\begin{align}
\beta^{\prime \prime} - \gamma^{\prime \prime} - \left(\beta^\prime
- \gamma^\prime\right)\left(\alpha^\prime - \tau^\prime - \frac{\bar
v_0^0}{2c_1}\, e^{\alpha - \tau}\right) &= 0,
\label{00-22a}\\
\mu^{\prime \prime} - \gamma^{\prime \prime} - \left(\mu^\prime -
\gamma^\prime\right)\left(\alpha^\prime - \tau^\prime - \frac{\bar
v_0^0}{2c_2}\, e^{\alpha - \tau}\right) &= 0, \label{00-33a}
\end{align}
\end{subequations}
with the solutions

\begin{align}
\beta^\prime &= \gamma^\prime + N_1 e^{\alpha - \tau - (\bar
v_0^0/2c_1)\int e^{(\alpha - \tau)}du}, \label{betares}\\
\mu^\prime &= \gamma^\prime + N_2 e^{\alpha - \tau - (\bar
v_0^0/2c_2)\int e^{(\alpha - \tau)}du}, \label{mures}
\end{align}

with $N_1$ and $N_2$ being the constants.

Then on account of \eqref{tau} we find

\begin{equation}
\gamma^\prime = \frac{1}{3}\left[\tau^\prime - e^{\alpha - \tau}
\left(N_1 e^{- (\bar v_0^0/2c_1)\int e^{(\alpha - \tau)}du}+ N_2 e^{
- (\bar v_0^0/2c_2)\int e^{(\alpha - \tau)}du}\right)\right].
\label{gam}
\end{equation}

Note that \eqref{EE10} leads to $\bar v^1_0 = 0$ while \eqref{EE23}
allows two possibilities: (i) $A^0_0 = 0$ and/or (ii)
$\left(\beta^\prime - \mu^\prime\right) = 0$. As one sees from
\eqref{bet0a} and \eqref{mu0a} the second condition is valid only if
$c_1 = c_2$ and $N_1 = N_2$ in the above equations.

Note that to determine the metric functions $\gamma$, $\beta$ and
$\mu$ we have to derive the equation for $\tau$. Summation of
\eqref{EE22}-\eqref{EE00} and 3 times \eqref{EE11} gives the
equation for $\tau$. Hence finally we obtain the system of equations
for defining the metric functions that takes into account the
restrictions those occur due to nontrivial non-diagonal components
of the Einstein equations.

\begin{subequations}
\label{sys0}
\begin{align}
\tau^{\prime \prime} + \tau^{\prime 2} - \tau^\prime \alpha^\prime
&= \frac{\kappa}{2} e^{2\alpha}\left[3 m_{\rm sp} S_0 e^{-\tau} + 3
\lambda \left(2 F - K F_K\right) - 2 \omega \bar v^0_0 e^{-\gamma -
\tau}\right], \label{tau1}\\
\gamma^\prime &= \frac{1}{3}\left[\tau^\prime - e^{\alpha - \tau}
\left(N_1 e^{- c_1^0\int e^{(\alpha - \tau)}du}+ N_2 e^{ - c_2^0\int
e^{(\alpha - \tau)}du}\right)\right],
\label{gam1}\\
\beta^\prime &= \frac{1}{3}\left[\tau^\prime + e^{\alpha - \tau}
\left(2 N_1 e^{- _1^0\int e^{(\alpha - \tau)}du} - N_2 e^{ - c_2^0
\int e^{(\alpha - \tau)}du}\right)\right],
\label{beta1}\\
\mu^\prime &= \frac{1}{3}\left[\tau^\prime + e^{\alpha - \tau}
\left(-N_1 e^{- c_1^0\int e^{(\alpha - \tau)}du}+ 2 N_2 e^{ - c_2^0
\int e^{(\alpha - \tau)}du}\right)\right]. \label{mu1}
\end{align}
\end{subequations}

As one sees, here we don't have any equation for defining $\alpha$.
So we need some additional conditions. We will solve the equations
for determining the metric functions \eqref{bet0a}, \eqref{mu0a},
\eqref{gam} and \eqref{tau1} numerically. The nonlinear term $F(K)$
we choose in the form $F(K) = K^n$ that gives $\left(2 F - K
F_K\right) = (2 - n) K^n = (2 - n) K_0^n e^{-2n\tau}$.

Further we consider two cases.

{Case 1.} In what follows we consider the harmonic radial coordinate
$u$. In this case the following condition holds for the metric
functions \cite{bron}
\begin{equation}
\alpha = \gamma + \beta + \mu. \label{hc}
\end{equation}

Taking into account that in this case $\tau = \alpha$ for a more
general solution to the Einstein equations with massive and
nonlinear spinor field as source we rewrite it in the form of
Cauchy:

\begin{subequations}
\label{sys1}
\begin{align}
\tau^\prime &= \eta, \label{taus1}\\
\eta^\prime &= \frac{\kappa}{2}e^{2\tau}\,\left[3 m_{\rm sp} S_0
e^{-\tau} + 3 \lambda (2 - n) K_0^n e^{-2n\tau} - 2 \omega \bar
v^0_0 e^{-\gamma - \tau}\right], \label{etas1}\\
\gamma^\prime &= \frac{1}{3}\left[\eta - \left(N_1 e^{- c_1^0 u}+
N_2 e^{- c_2^0 u}\right)\right], \label{gams1}\\
\beta^\prime &= \frac{1}{3}\left[\eta + \left(2 N_1 e^{-
c_1^0 u} - N_2 e^{- c_2^0 u}\right) \right], \label{bets1}\\
\mu^\prime &= \frac{1}{3}\left[\eta + \left(- N_1 e^{- c_1^0 u} +
N_2 e^{- c_2^0 u}\right) \right]. \label{mus1}
\end{align}
\end{subequations}

Though the choice of harmonic radial coordinate given by \eqref{hc}
essentially simplifies the initial system \eqref{sys0}, nevertheless
it is hard to find exact solutions to the system \eqref{sys1}. That
is why we solve the system \eqref{sys1} numerically. Right now our
goal is to exhibit some qualitative solutions. Taking it into
account we modify the system of equations as simple as possible. In
doing so in \eqref{sys1} and in \eqref{sys2} that will follow later,
we set $S_0 = 1,\, K_0 = 1,\, \bar v^0_0 = 1$. We also set $c_1^0 =
1,\, c_2^0 = 2,\, N_1 = 2,\, N_2 = 3,\, m_{\rm sp} = 1,\, \omega =
1$ and $\lambda = 1$. The values the corresponding functions on the
axis of symmetry, i.e. at $u = 0$ are set to be trivial: $\tau_0 =
0,\,\eta_0 = 0,\, \gamma_0 = 0,\, \beta_0 = 0, \, \mu_0 = 0$. We
consider two different values of $n$. In case of $n = 5$ the
nonlinear term takes the form $-e^{-10\tau}$, while for $n = -2$ the
nonlinear term becomes $2 e^{4\tau}$.

As one can see from \eqref{gams1}, \eqref{bets1} and \eqref{mus1}
depending on the value of $c_1$ and $c_2$ they might vary relative
to $\tau$, whereas the behavior of $\tau$ depends on not only the
choice of spinor field nonlinearity, but also on the choice of
coordinate condition. In Fig. \ref{Fig1} and Fig. \ref{Fig2} we
illustrate the behavior of the metric functions for the spinor field
nonlinearity being a power law of invariant $K$. While the Fig.
\ref{Fig1} corresponds to the case with $n =5$, Fig. \ref{Fig2}
corresponds to the case with $n=-2$.

\myfigures{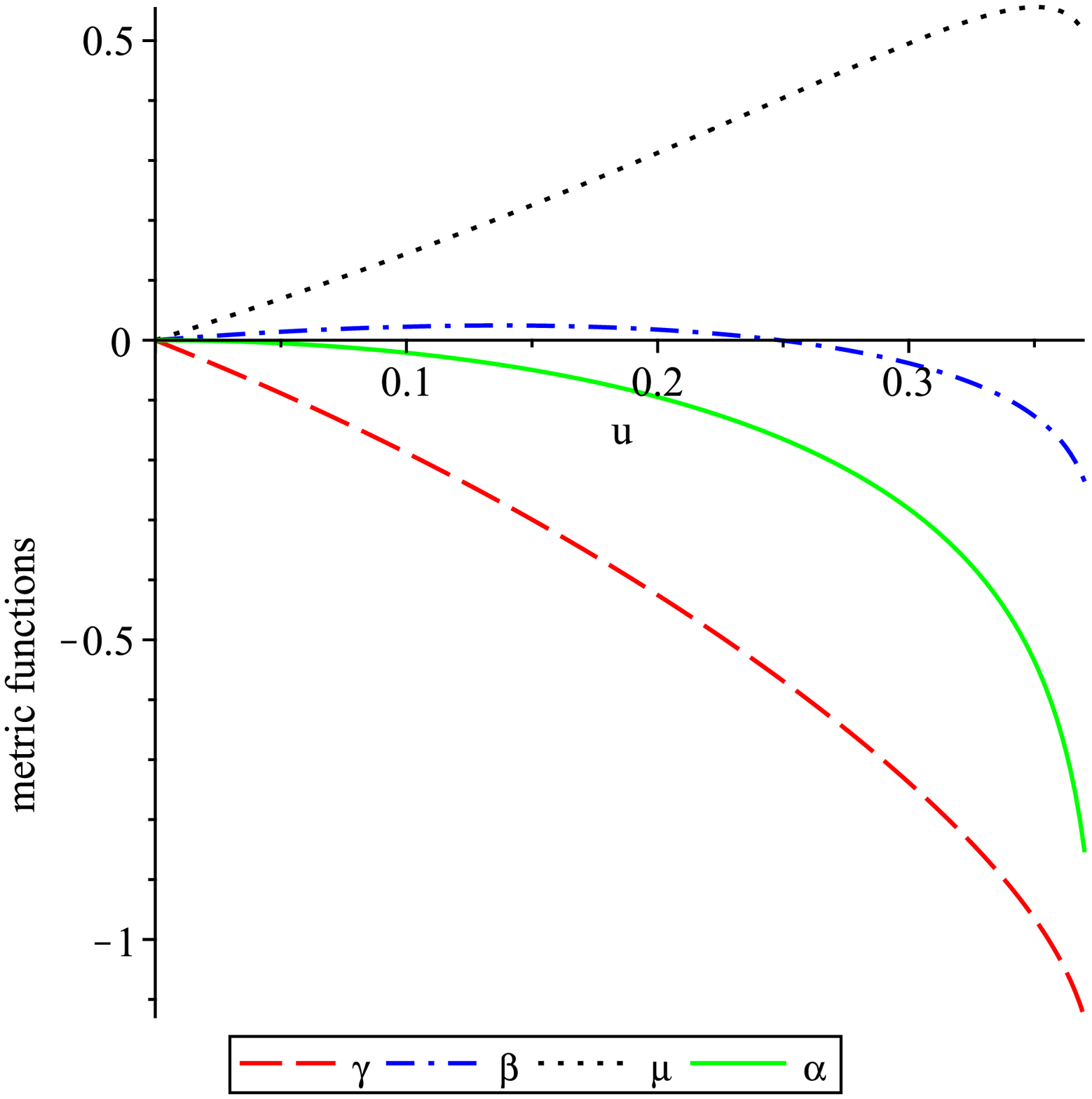}{0.40}{Evolution of the metric functions for $n =
5$}{0.45}{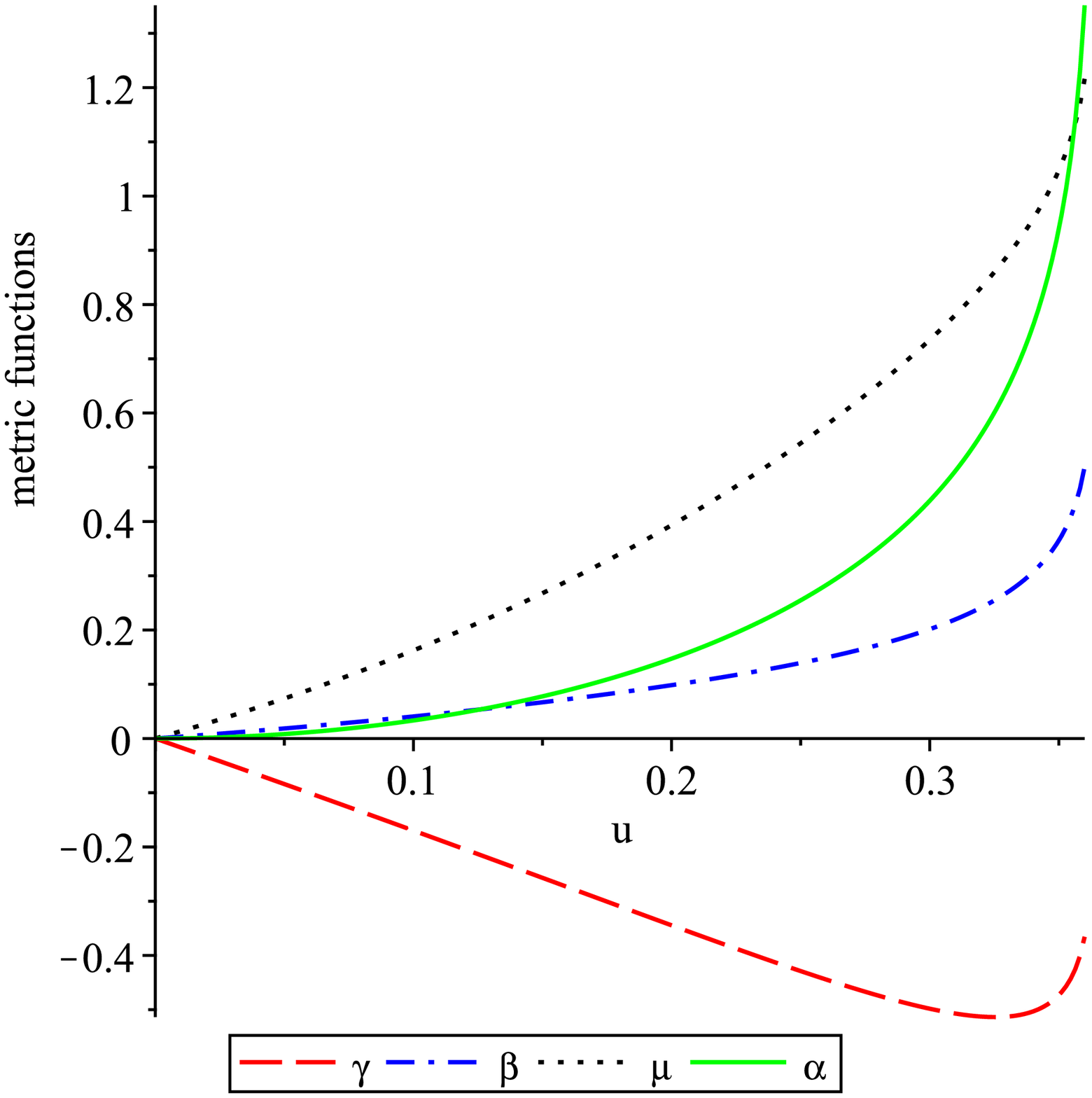}{0.40}{Evolution of the metric functions for
$n=-2$}{0.45}

{Case 2:} As a second choice let us consider the quasiblogal
coordinate $\alpha = - \gamma$ \cite{BronBook}. In this case we have

\begin{subequations}
\label{sys2}
\begin{align}
\tau^\prime &= \eta, \label{taus2}\\
\eta^\prime &= - \frac{4}{3} \eta^2 + \frac{1}{3} \eta e^{-\gamma -
\tau} \left(N_1 e^{- c_1^0 \int e^{(-\gamma - \tau)}du}+
N_2 e^{- c_2^0 \int e^{(-\gamma - \tau)}du}\right)  \nonumber \\
&+ \frac{\kappa}{2}e^{-2\gamma}\,\left[3 m_{\rm sp} S_0 e^{-\tau} +
3 \lambda (2 - n) K_0^n e^{-2n\tau} - 2 \omega \bar v^0_0 e^{-\gamma
- \tau}\right], \label{etas1}\\
\gamma^\prime &= \frac{1}{3}\left[\eta - e^{-\gamma - \tau}
\left(N_1 e^{- c_1^0 \int e^{(-\gamma - \tau)}du}+ N_2
e^{- c_2^0 \int e^{(-\gamma - \tau)}du}\right)\right], \label{gams2}\\
\beta^\prime &= \frac{1}{3}\left[\eta + e^{-\gamma - \tau} \left(2
N_1 e^{- c_1^0 \int e^{(-\gamma - \tau)}du} - N_2
e^{ - c_2^0 \int e^{(-\gamma - \tau)}du}\right)\right], \label{bets2}\\
\mu^\prime &= \frac{1}{3}\left[\eta + e^{-\gamma - \tau} \left(-N_1
e^{- c_1^0 \int e^{(-\gamma - \tau)}du}+ 2 N_2 e^{ - c_2^0 \int
e^{(-\gamma - \tau)}du}\right)\right] \label{mus2}
\end{align}
\end{subequations}

As in previous case the system \eqref{sys2} is also solved
numerically. As was mentioned earlier, like the foregoing case we
set $S_0 = 1,\, K_0 = 1,\, \bar v^0_0 = 1$. We also set $c_1^0 =
1,\, c_2^0 = 2,\, N_1 = 2,\, N_2 = 3,\, m_{\rm sp} = 1,\, \omega =
1$ and $\lambda = 1$. Analogical to the previous case we consider
that the values the corresponding functions on the axis of symmetry,
i.e. at $u = 0$ they are trivial: $\tau_0 = 0,\,\eta_0 = 0,\,
\gamma_0 = 0,\, \beta_0 = 0, \, \mu_0 = 0$. We consider two
different values of $n$. In case of $n = 5$ the nonlinear term takes
the form $-e^{-10\tau}$, while for $n = -2$ the nonlinear term
becomes $2 e^{4\tau}$. In Fig. \ref{Fig3} we draw the metric
functions those correspond to the nonlinear term being a power law
of invariant $K$ with $n =5$, while Fig. \ref{Fig4} corresponds to
the case with $n=-2$.

\myfigures{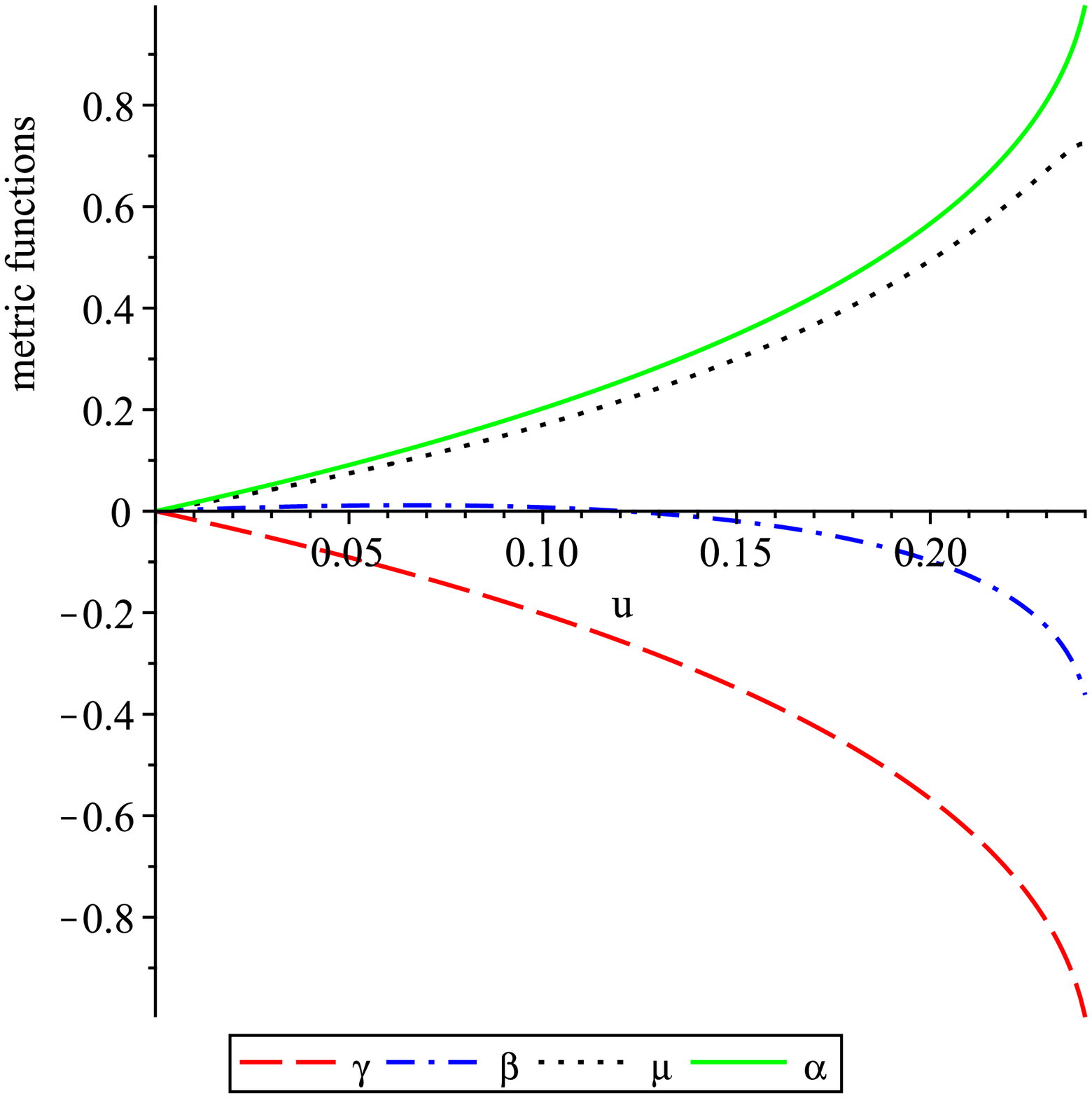}{0.40}{Evolution of the metric functions for
$n=5$}{0.45}{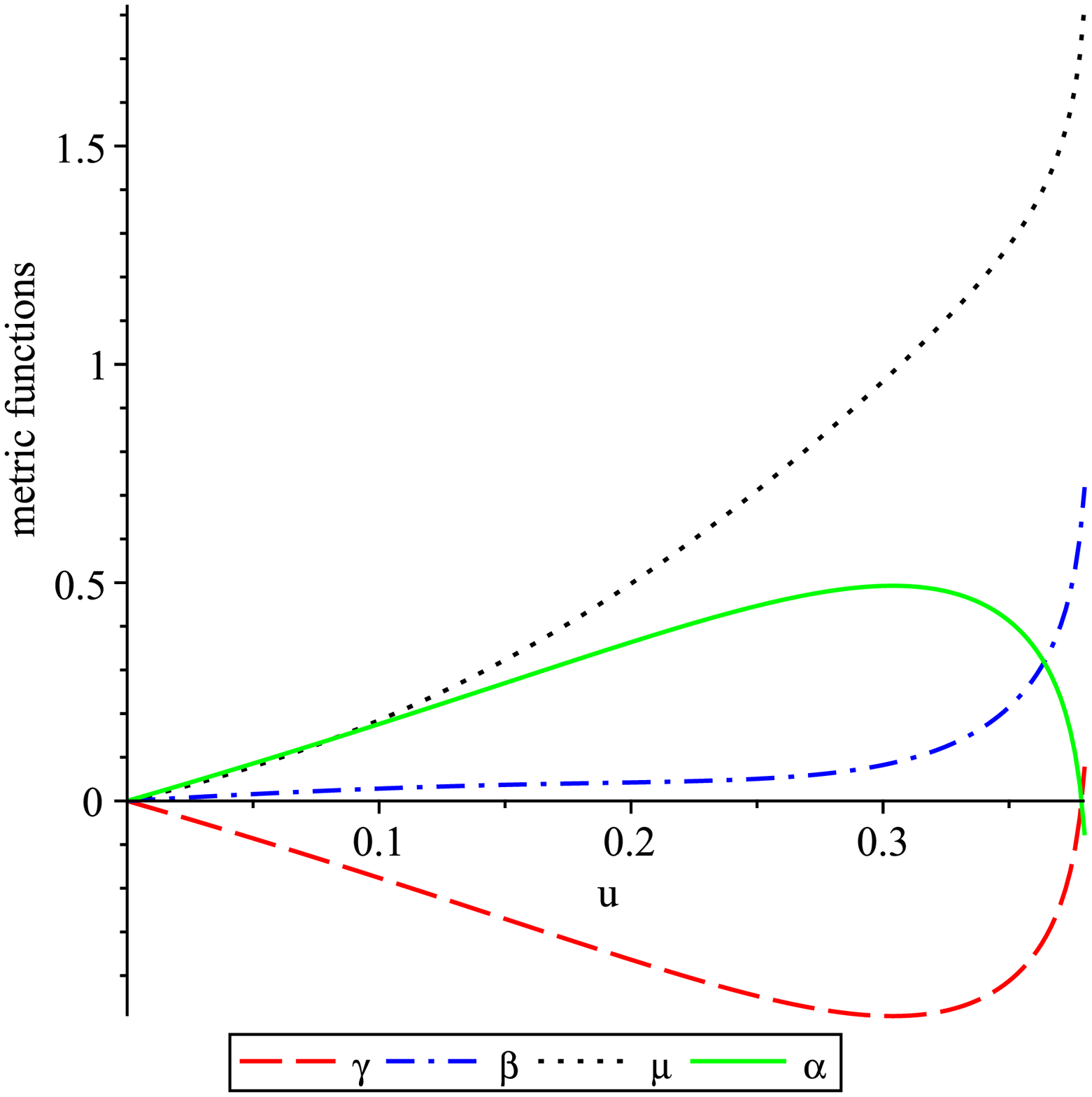}{0.40}{Evolution of the metric functions for $n =
-2$}{0.45}

\section{Conclusions}

We studied a system of nonlinear spinor field minimally coupled to a
static cylindrically symmetric space-time. We consider the spinor
field that depends on time and radial coordinates. The time
dependence is taken in such a way that the invariants of the spinor
field remain time independent. The energy-momentum tensor (EMT) of
the spinor field possesses nontrivial non-diagonal components. While
in case of a static spinor field the presence of non-diagonal
components of the EMT imposes three-way restrictions on the
space-time geometry and/or on the components of the spinor field
similar to the Bianchi type-I cosmological model \cite{universe}, in
case of a time dependent spinor field things become more complicate.
Moreover, while in a static spherically symmetric space-time the
presence of non-trivial non-diagonal components of EMT of the spinor
field has no effect on the space-time geometry \cite{SpinSphBRS}, in
static cylindrically symmetric space-time it influences both the
space-time geometry and the components of the spinor field. Unlike
the static case in the present model we have $T_0^0 \ne T_2^2$.  It
should be noted that the expressions $(T_0^0 + T_1^1)$ and $(T_0^0 -
T_1^1)$ can be both positive and negative, depending on the type of
nonlinearity.

\vskip 1 cm

\noindent
\textbf{DAS:} No datasets were generated or analysed
during the current study


\end{document}